\documentclass{elsart}

\usepackage{amssymb}
\usepackage{graphicx}

\def\la{\lambda}
\def\lb{\frac{\Lambda}{3}}
\def\La{\Lambda}
\def\th{\theta}
\def\vp{\varphi}
\def\a{\alpha}
\def\p{\varrho}
\def\s{\sigma}

\def\ra{\rightarrow}

\begin{document}

\begin{frontmatter}



\title{Rotating De Sitter Space}

\author[label1]{G. F. Chapline}
\author[label2]{P. Marecki}

 \address[label1]{Lawrence Livermore National Laboratory, Livermore, CA 94550}
 \address[label2]{Institut fuer Theoretische Physik, D-04009 Leipzig, Germany}

\author{}

\address{}

\begin{abstract}

An exact solution of the vacuum Einstein equations with a cosmological
constant is exhibited which can perhaps be used to describe the interior of
compact rotating objects. The physical part of this solution has the
topology of a torus, which may shed light on the origin of highly
collimated jets from compact objects.
\end{abstract}

\end{frontmatter}

\section{Introduction}

In this paper we exhibit an exact solution to the Einstein field
equations that may help resolve two outstanding puzzles in theoretical
astrophysics. The first puzzle is to describe the nature of space-time inside a
rotating object that is sufficiently compact that it lies entirely inside a
surface where classical general relativity predicts that an event horizon
would form. The conventional view is that such an object is a ``black hole''.
However, both non-rotating and rotating black holes have features such as
singularities and ``reversal of space and time'' that may be unphysical. In
addition, the Kerr solution for a rotating black hole \cite{Kerr} shares in common
with other rotating solutions of Einstein's equations the pathological feature
that there are closed time-like curves. It has been pointed out \cite{CHHLS,MM} that in the
case of non-rotating compact objects the objectionable features of the non-rotating black
hole interior space-time would be removed if the interior Schwarzchild
space-time were replaced with de Sitter's ``interior'' cosmological solution
\cite{deSitter}. This space is non-singular and removes the ``reversal of space and time'' that plagues the interior Schwarzschild solution. Furthermore the de Sitter
interior solution can be made to exactly match the exterior Schwarzschild
solution at the event horizon if the vacuum energy is chosen so that the total
mass-energy of the interior de Sitter solution matches the black hole mass.
According to this new picture of non-rotating compact objects space-time
would not be analytically smooth at the event horizon, so classical general
relativity would fail there. However, it has long been recognized that
quantum effects become important near an event horizon, and therefore it is
quite plausible that classical general relativity fails in the vicinity of an event
horizon. References \cite{CHHLS} and \cite{MM} offer two different scenarios as to what actually happens at the event horizon. However, for the purposes of this paper it not
necessary to understand in detail what happens at the event horizon; instead
we will focus on question as to whether there is a candidate space-time that
could serve as a non-singular model for the bulk interior space-time inside
rotating compact objects.

In accordance with the expectation that the interior space-time of a collapsed object should be
obtained by continuous ``squeezing'' of a condensate vacuum state \cite{Chapline}, we
expect that this space-time should have a large vacuum energy; i.e. this interior space-time should locally resemble classical flat space-time with a cosmological constant; i.e. it should locally look like a region of de Sitter space-time. One nagging question concerning
the proposals of references \cite{CHHLS} and \cite{MM}, though, is what should replace de
Sitter's interior solution in the case of a rotating compact object. To our
knowledge a rotating version of de Sitter space-time has never been
explicitly discussed in the literature. In the following we address this
deficiency by exhibiting a mathematically exact solution of Einstein's field
equations that is in fact a rotating generalization of de Sitter's interior
solution. Not only does this solution provide a plausible picture for the
nature of space-time in the interior of rotating compact objects, but as a
bonus this solution provides a new insight into the nature of the highly
collimated jets that have been observed to be emanating from compact astrophysical objects.

\newpage
\section{The metric}

Our proposed interior metric is (we use units such that $8\pi G/c^2=1$)
\begin{eqnarray}
ds^2&=\left[1-\lb(r^2-a^2\cos^2\th )\right]dt^2+2a \left[1- \lb r^2\cos^2\th\right]dtd\vp \nonumber \\
&-\frac{r^2+a^2\cos^2\th}{r^2-\lb r^4+a^2}\, dr^2-\frac{r^2+a^2\cos^2\th}{1-\lb a^2 \cos^2\th \cot^2\th}\, d\th^2\label{metric}\\&-(r^2\sin^2\th-a^2\cos^2\th )d\vp^2\nonumber
\end{eqnarray}
where $a$ is the angular momentum per unit mass. This metric is a limiting
case of a class of metrics discovered by Carter \cite{Carter} and independently by
Plebanski \cite{Plebanski}. In the limit $a \ra 0$ the metric (\ref{metric}) reduces to de Sitter's 1917 metric. It is clear by inspection that in contrast with the interior Kerr metric
there are no space-time singularities near to $r=0$ for any value of $\th$. The
apparent singularities in the $g_{rr}$ and $g_{\th\th}$
components of the metric tensor can be removed by a change of variables \cite{Plebanski}, and represent event horizons where \mbox{$g_{00}g_{\vp\vp}-g_{0\vp}^2= 0$}. The singularity in $g_{rr}$
is associated with a spherical event
horizon located at 
\begin{equation}\label{horizon}
\hspace{3cm} r_H^2=\frac{3}{2\La}+\left[\frac{9}{4\La^2}+\frac{3a^2}{\La}\right]^{1/2}
\end{equation}

In the limit $a\ra 0\,$ $r_H$
becomes the de Sitter horizon $\sqrt{3/\La}$. In addition to the
spherical event horizon (\ref{horizon}) there is a conical event horizon located at
\begin{equation}\label{ch}
\hspace{3cm}  \tan^2\th_H=-\frac{1}{2}+\sqrt{\lb a^2+\frac{1}{4}}
\end{equation}
In the case of slow rotation $\La a\ll 1$ this conical event horizon is located
very near to the axis of rotation.

The nemesis of rotating space-times, closed time-like curves, will
appear if $g_{\vp \vp}$ is positive for some values of $r$. In our case this means that
\begin{equation}\label{ctc}
\hspace{3cm}  r^2\sin^2\th-a^2\cos^2\th<0
\end{equation}
This is will be satisfied if $\p<a$, where $\p$ is the horizontal distance from the
axis of rotation. Thus for slow rotation closed time-like curves will appear
very close to the rotation axis. This is very reminiscent of the situation with
space-time spinning strings \cite{Mazur}. Actually for all values of the rotation parameter $a$ the
conical horizon (\ref{ch}) lies inside the region where closed time-like curves
appear, and the critical angle where the inequality in (\ref{ctc}) becomes an equality
is precisely the horizon angle $\th_H$ where $r=r_H$.

The event horizon for the Kerr solution will match the event horizon
(\ref{horizon}) if the mass parameter for the Kerr solution is
\begin{equation}\label{mass_cond}
\hspace{3cm}   m=\frac{\La}{6}\, r_H^3
\end{equation}

Curiously this is the same condition that was used in ref. \cite{CHHLS} to match de
Sitter's interior solution to the exterior Schwarzschild solution in the case of
a non-rotating compact object. If we impose the condition (\ref{mass_cond}) the event
horizon for our ``interior'' solution will occur at precisely the same radius as
the horizon for the Kerr solution. In this case it might be reasonable to
suppose that the ``exterior'' space-time outside the horizon (\ref{horizon}) is just the
usual exterior Kerr solution. Near to the spherical event horizon (\ref{horizon}) the
angular part of our rotating metric (\ref{metric}) has the form
\begin{equation}
\hspace{-0.6cm}  ds^2=-a^2\, \frac{\sin^2\th-\lb a^2\cos^4\th}{r_H^2+a^2\cos^2\th}\left(dt-\frac{r^2_H}{a}d\vp\right)^2-\frac{r_H^2+a^2\cos^2\th}{1-\lb\,a^2\cos^2\th\cot^2\th}\, d\th^2.
\end{equation}

For comparison the angular part of the Kerr metric when expressed in
Boyer-Lindquist coordinates \cite{BL} and evaluated on the event horizon is
\begin{equation}
ds^2=-a^2\,\frac{\sin^2\th}{r_H^2+a^2\cos^2\th}\left(dt-\frac{r_H^2}{a}d\vp\right)^2-(r_H^2+a^2\cos^2\th)d\th^2.
\end{equation}
It can be seen that except near to $\th=0$ the angular part of our metric near 
to the spherical event horizon is not too different to the angular part of the
Kerr metric at the event horizon for all values of $a$ such that $\La a^2<1$.
Significant difference do appear near to $\th=0$, which as we discuss below is
due to the appearance of new physics near to the axis of rotation.

Inside the spherical event horizon (\ref{horizon}) the behavior of our metric is
completely different from that of the interior Kerr metric. For example, there
are no space-time singularities. In the case of the Kerr solution $g_{00}<0$
everywhere inside the ``ergosphere'' whose outer boundary lies outside the
event horizon at $r^2+a^2\cos^2\th = 2mr$. Although our $g_{00}$ is negative at the event
horizon (and close to the Kerr $g_{00}$), it is actually positive for all values of $r$
inside $r^2= (3/\La)^{1/2}+a^2\cos^2\th$, which for small $\La a^2$
would be almost everywhere in the interior. In addition in contrast with the Kerr solution the radial metric coefficient $g_{rr}$ is negative for all values of $r$ inside the spherical
event horizon. Thus the problematic reversal of the roles of time and radial
distance in the interior Kerr solution is alleviated.
The Kerr solution has the property that inside the ergosphere particles
cannot be at rest but must rotate about the axis. At the event horizon the
frame in which particles could be at rest rotates with the ``frame dragging''
angular velocity
\begin{equation}
\hspace{3cm}  \left. \frac{d\vp}{dt}\right|_{r=r_H}=\frac{a}{r_H^2+a^2}.
\end{equation}

 For the metric (\ref{metric}) $g_{00}<0$ at the spherical event horizon and the frame
rotation velocity is $a/r_H^2$, so particles in our space-time will also rotate as
they approach the event horizon from the inside. Indeed our interior metric
contains a reflection of the usual exterior Kerr ergosphere, with an inside
boundary at $r^2= (3/\La)^{1/2}+ a^2 \cos^2\th$. Thus in our picture of rotating compact
objects the metric just inside the event horizon is a reflection of the metric
just outside, at least away from $\th=0$. Reflection symmetry between the
inner and outer metrics at an event horizon is just the matching condition for
metrics suggested in ref. \cite{CHHLS}, and is a consequence of replacing the smooth
geometry at an event horizon that is predicted by classical general relativity
with a quantum critical layer.

\section{Comparison with Demianski and Plebanski metrics}
One might guess that the metric (\ref{metric}) could also be derived from
Demianski's well known generalization of the Kerr solution to include a
cosmological constant \cite{Demianski}. Indeed taking the $m=0$ limit of Demianski's
metric yields
\begin{eqnarray}
 ds^2&=&\left[1-\la (r^2+a^2\sin^2\th )\right]dt^2+2a\sin^2\th\cdot \la (r^2+a^2)\, dtd\vp\nonumber\\
&-&\frac{r^2+a^2\cos^2\th}{(r^2+a^2)(1-\la r^2)}\, dr^2-\frac{r^2+a^2\cos^2\th}{1+\la a^2 \cos^2\th}\, d\th^2\label{dem_metric}\\
&-&(a^2+r^2)\sin^2\th\left[1+\la a^2\right]d\vp^2,\nonumber
\end{eqnarray}
where $\la=\frac{\La}{3}$. As was the case for metric (\ref{metric}) the variables $\th$ and $\vp$ represent the polar
angles on a sphere. At first sight (\ref{metric}) and (\ref{dem_metric}) appear to be different. However, both metrics (\ref{metric}) and (\ref{dem_metric}) can be obtained as special cases of the Plebanski metric \cite{Plebanski}, which has the general form
\begin{eqnarray}
 ds^2&=&\frac{\mathcal Q}{p^2+q^2}\, (dt-p^2d\s)^2-\frac{\mathcal P}{p^2+q^2}\, (dt+q^2d\s)^2\nonumber\\&-&
\frac{p^2+q^2}{\mathcal P}\, dp^2-\frac{p^2+q^2}{\mathcal Q}\, dq^2\label{pleb_metric}
\end{eqnarray}
When the mass, NUT charge, and electric and magnetic charges are all zero
then the functions $\mathcal P(p)$ and $\mathcal Q(q)$ have the simple forms
$\mathcal P=b-\epsilon p^2-\la p^4$ and $\mathcal Q=b+\epsilon q^2-\la q^4$. In this case the Weyl tensor vanishes and the Plebanski metrics are conformally flat. For both metrics (\ref{metric}) and (\ref{dem_metric}) $p=a\cos\th$, $q = r$, $\s=\vp/a$, and
$b = a^2$. However, for metric (\ref{metric}) $\epsilon=1$ and $\tau= t$, while for the metric (\ref{dem_metric}) $\epsilon=1-\la a^3$ and $\tau=t + a^2\s$.

Evidently the essential difference between the two geometries lies in
the value of $\epsilon$. However it is known that Plebanski metrics with different
values of are related by a certain scaling transformation. This scaling
transformation has the form
\begin{equation}\label{scaling}
p' = p/\a,\ q' = q/\a,\ \s'=\a^3\s,\ \tau'=\a\tau,\ b'=b/\a^4,\ \epsilon'=\epsilon/\a^2,
\end{equation}
where $\a$ is the scaling parameter. The value of $\La$ is unchanged. Because
$b = a^2$ for both metrics we may replace $a$ by $\sqrt{b}$ so that both metrics depend only on the parameters $b$ that appear in the original Plebanski metric (\ref{pleb_metric}).
That the metrics (\ref{metric}) and (\ref{dem_metric}) are locally isometric can now be seen as follows: we start with the $m=0$ Demianski metric (\ref{dem_metric}) and rescale it using the
scaling transformation (\ref{scaling}) and $\a=1-\la a^2$. This leads to the metric (\ref{metric}) with $\epsilon=1$ and $a^2=b/(1-\La b/3)$. Therefore the $m=0$ Demianski's geometry is locally isometric to the geometry of (\ref{metric}).
The question to whether the $m=0$ Demianski geometry is isomorphic
to our rotating solution is more complicated because the value of $b$ affects
the ranges of $p$ and $\vp$. In particular since $\vp= 0$ is identified with
$\vp= 2\pi$, then in the scaled metric $\s= 0$ is identified with $\s= 2\pi / \sqrt{b}$. In addition the range of $p$ is restricted because $p\in[ \sqrt b,+ \sqrt b]$. Thus whereas the metric (\ref{metric}) is defined
for all values of $\th$ and $\vp$ the $m=0$ Demianski geometry corresponds to only a
part of the sphere (cf. Fig. 1). Amusingly the latitudes covered by the $m=0$ Demianski
are just those outside the conical horizon eq. (\ref{ch}).
In accordance with our a priori expectations
regarding the nature of the vacuum state inside
compact objects, both metrics are locally
isometric to de Sitter space-time.
\begin{figure}
\begin{center}
\includegraphics[scale=0.75]{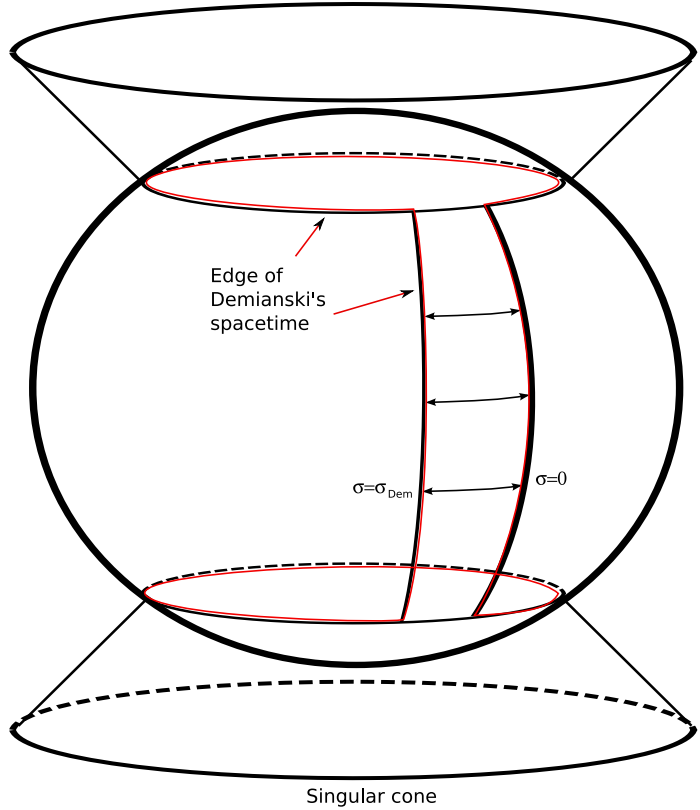}
\end{center}
\caption{The region of the spacetime (\ref{metric}) isometric (covered) by the Demianski's spacetime (\ref{dem_metric}). 
}
\end{figure}

\section{Behavior near to the axis of rotation}
As noted above the metric (\ref{metric}) is plagued by time-like closed curves
near to the axis of rotation. Closed time-like curves are extremely
pathological from the point of view of quantum mechanics. The pathological
nature of the space-time corresponding to the metric (\ref{metric}) near to the axis of rotation can also be seen from the signature of the metric. Physical space-times should have the signature $+\, -\, -\, -$. In the case of the Plebanski metrics
this is only possible if $\mathcal P > 0$, $\mathcal Q > 0$ or $\mathcal P > 0$, $\mathcal Q < 0$. If $\mathcal P < 0$, as is the case
inside he conical horizon, the signature is $+\, +\, +\, -$, so the space-time is not
physical. Both of these considerations suggest that space-time undergoes
some sort of phase transition near to the axis of rotation.

Recently it has been suggested \cite{CHM04} that the way to resolve the
difficulties with classical rotating space-times that have closed time-like
curves is to suppose that the rotation is actually carried by space-time
``spinning strings'' \cite{Mazur}, in a manner analogous to the way rotation of
superfluid helium inside a rotating container is carried by quantized vortices.
The spinning strings resolve the question of the consistency of rotating
space-times with quantum mechanics because the vorticity of space-timewould be concentrated into the cores of the spinning strings where the
condensate density would be very low and the Einstein equations are
modified by the appearence of torsion. As shown in ref. \cite{CHM04} averaging over
the vorticity of many perfectly aligned spinning strings leads to a Godel-like
space-time. In a similar way it is reasonable to guess that the correct
physical picture for the space-time inside the region where closed time-like
curves appear in our solution for a rotating compact object is a Godel-like
space-time. Indeed the Som-Raychaudhuri metric exhibited in ref. \cite{CHM04} may
be a good approximation for the metric in this region. This metric would be
applicable inside the critical radius where the local speed of frame rotation
is equal to the speed of light.
The equation of motion for particles in a Godel-like space-time is well
known \cite{Kundt}. In general this flow of particles will be collimated since the
particles are confined to lie inside the cylinder where the
velocity of frame rotation is less than the speed
of light. In our situation the radius of this
cylinder will equal the angular momentum per unit
mass parameter a used in eq. (\ref{metric}); i.e. where the
closed time-like curves first appear in our
solution as the axis of rotation is approached.  On the other hand particles in a Godel-like
space-time are free to move parallel to the axis, so that for slow rotation of
our compact object the flow of particles along the axis of rotation will be
highly collimated.

\section{Summary}

In summary, the metric (\ref{metric}) provides an interior solution for rotating
compact objects that avoids many of the unphysical features of the interior
Kerr solution. It does not avoid the appearance of closed time-like curves,
but the results of ref. \cite{CHM04} suggest that near to the axis of rotation a Godel-like
phase of space-time appears where the vacuum energy is much smaller than
in the bulk condensate of the rotating object and solid body-like rotation of
the space-time appears.

Finally we should note that our metric (\ref{metric}) may also serve as a model for the
large scale structure of our universe, where there are hints from observations of the large scale anisotropy of the cosmic microwave
background that the universe might be rotating \cite{CH06}.

The authors are very grateful for numerous
discussions with Pawel Mazur.



\end{document}